# Crystalline-dependent magnon torques in all-sputtered Hf/Cr$_2$O$_3$/ferromagnet heterostructures

Yuchen Pu, Guoyi Shi, Chenhui Zhang, Xinhou Chen, Hanbum Park, and Hyunsoo Yang*

Department of Electrical and Computer Engineering, National University of Singapore, Singapore 117583, Singapore

E-mail: eleyang@nus.edu.sg

Electron motion in spin-orbit torque devices inevitably leads to the Joule heating issue. Magnon torques can potentially circumvent this issue, as it enables the transport of spin angular momentum in insulating magnetic materials. In this work, we fabricate a sandwich structure composed of Hf/antiferromagnetic Cr$_2$O$_3$/ferromagnet and demonstrate that the magnon torque is strongly dependent on the crystalline structure of Cr$_2$O$_3$. Magnon torques are stronger when the Néel vector of Cr$_2$O$_3$ aligns parallel to the spin polarization generated in Hf, while they are suppressed when the Néel vector is perpendicular to the spin polarization. The magnon torque efficiency is estimated to be −0.134 using in-plane second harmonic Hall measurements. Using magnon torques, we achieve perpendicular magnetization switching of CoFeB, with a critical switching current density of $4.09 \times 10^7$ A/cm$^2$. Furthermore, the spin angular momentum loss due to the insertion of Cr$_2$O$_3$ is found to be lower than that of polycrystalline NiO. Our work highlights the role of antiferromagnet crystalline structures in controlling magnon torques, broadening the potential applications of magnon torques.

## 1. Introduction

Spin-orbit torque (SOT) is an efficient and fast mechanism for manipulating the magnetization of magnetic materials, making it a promising candidate for application in logic and magnetic memory devices.[1] In SOT devices, spin currents, generated from the spin source layer with spin-orbit interaction, exert torques on the adjacent ferromagnetic layer to achieve magnetization switching.[2] However, electron-mediated spin torque, which involves the motion of electrons, inevitably leads to Joule heating.[1c] In integrated circuits with a high packing device density, this Joule heating causes significant energy loss and degrades overall performance. Moreover, the spin propagation length of spin currents is typically on the order of nanometers,[3] preventing the transmission of spin information over long distances.

Magnons, the quanta of spin waves, can carry spin angular momentum and propagate in insulating systems, potentially circumventing this Joule heating issue.[4] Magnon currents can generate magnon torques in a ferromagnet (FM), modulating its magnetization. Therefore, the use of magnon torques is an alternative approach for manipulating the magnetization with less energy loss. Magnons also have several advantages, including micrometer-scale transmission length[5] and ultrafast propagation velocity,[6] paving the way for the development of high-speed and low-power-consumption spintronic devices. Moreover, the spin source and ferromagnetic layer can be electrically isolated by antiferromagnetic insulator, which may relax the modern interconnect requirements in integrated circuit (IC) design.

In previous studies, magnetization switching using magnon torques in a FM layer with in-plane magnetic anisotropy has been demonstrated in trilayer structures composed of $Bi_2Se_3$/NiO/NiFe.[7] With an external assisting magnetic field, the switching of FM layer with perpendicular magnetic anisotropy (PMA) has been achieved.[8] Recently, deterministic, field-free switching of FM with PMA is realized in $WTe_2$/NiO/CoFeB heterostructures, using out-of-plane anti-damping magnon torques.[9] In these studies, polycrystalline NiO layers are utilized, enabling the transmission of both in-plane and out-of-plane spin angular momentum, which is beneficial for field-free switching. However, the relationship between magnon torques and the orientation of the Néel vector in antiferromagnetic insulators remains poorly understood and has not been thoroughly investigated. Moreover, the reliance on topological materials in these studies presents a challenge to integrate into an industry compatible process due to the usage of topological materials such as $Bi_2Se_3$, $Bi_2Te_3$, and $WTe_2$ to generate spin currents.

In this work, we observe crystalline-dependent magnon torques in antiferromagnetic $Cr_2O_3$ instead of polycrystalline NiO. We fabricate sandwich heterostructures of Hf/$Cr_2O_3$/CoFeB on single crystal $Al_2O_3$ (0001) (c-$Al_2O_3$) and $Al_2O_3$ ($11\bar{2}0$) (a-$Al_2O_3$) substrates, achieving in-plane and out-of-plane Néel vector, respectively, in the $Cr_2O_3$ layer. We demonstrate that in-plane spin angular momentum, generated by the heavy metal Hf, propagates through $Cr_2O_3$ with the in-plane Néel vector and exerts magnon torques in the CoFeB layer. However, when the Néel vector in $Cr_2O_3$ is out-of-plane, magnon torques are suppressed. This study confirms that only spins collinear to the Néel vector can generate magnon torques in antiferromagnetic insulators. Furthermore, magnon-torque-driven switching is observed in all-sputtered heterostructures without topological materials, highlighting the compatibility with modern microfabrication processes.

## 2. Results and Discussion

$Cr_2O_3$ is an antiferromagnetic insulator with a hexagonal crystal structure belonging to the $R\bar{3}c$ space group, as illustrated in Figure 1a. It exhibits uniaxial anisotropy with magnetic moments of $Cr^{3+}$ pointing antiparallel with each other along the [0001] axis.[10] Hf is selected as the spin source layer due to its hexagonal crystal structure and minimal lattice mismatch with $Cr_2O_3$. Using magnetron sputtering, we deposit 8 nm thick Hf and 20 nm thick $Cr_2O_3$ on c-$Al_2O_3$ and a-$Al_2O_3$ substrates (details are shown in the Experimental Section). The X-ray diffraction (XRD) pattern in Figure 1b shows the $Cr_2O_3$ (0006) peak, confirming that the Néel vector (**n**) in the c-$Al_2O_3$/Hf/$Cr_2O_3$ system is out-of-plane. Conversely, Figure 1c reveals the $Cr_2O_3$ (11$\bar{2}$0) peak in the a-$Al_2O_3$/Hf/$Cr_2O_3$ system, verifying the in-plane **n** in $Cr_2O_3$. Furthermore, XRD $\phi$-scans for a-$Al_2O_3$/Hf/$Cr_2O_3$ clearly show the textured $Cr_2O_3$ on a-$Al_2O_3$ with a relationship of $Cr_2O_3$ (11$\bar{2}$0) // $Al_2O_3$ (11$\bar{2}$0) and $Cr_2O_3$ [0001] // $Al_2O_3$ [0001] (details are shown in Section S1, Supporting Information).

When a charge current is applied along the x-axis, a spin current with the spin polarization along the y-axis ($\sigma_y$) is generated in the Hf layer via the spin Hall effect. In the $Cr_2O_3$ (0001) layer with **n** along the z-axis, magnon currents cannot be induced due to the orthogonality between $\sigma_y$ and **n**, as illustrated in Figure 1d. On the other hand, in the $Cr_2O_3$ (11$\bar{2}$0) layer with **n** along the y-axis, magnon currents are induced as $\sigma_y$ aligns parallel to **n**, as depicted in Figure 1e. These magnon currents exert magnon torques, switching the magnetization in the FM. To quantify the torque efficiency, we utilize in-plane second harmonic Hall measurements (details are shown in Section S2, Supporting Information).[11] A 6 nm thick Py ($Ni_{81}Fe_{19}$) is sputtered on top of the Hf/$Cr_2O_3$ stack. The antiferromagnetic (AFM) order of $Cr_2O_3$ film is first characterized via the exchange bias effect, which is associated with the exchange coupling between the $Cr_2O_3$ (11$\bar{2}$0) and Py layers.[12] Figure 2a shows the exchange bias field ($\mu_0 H_{EB}$) of a-$Al_2O_3$/Hf (8 nm)/$Cr_2O_3$ (20 nm)/Py at various temperatures. A negative $\mu_0 H_{EB}$ is observed below the blocking temperature ($T_B$) of 170 K, indicating the AFM order of $Cr_2O_3$ film under 170 K. Furthermore, we also measure the spin Hall magnetoresistance (SMR) of Hf (8 nm)/$Cr_2O_3$ (20 nm) bilayers, confirming a paramagnetic to antiferromagnetic transition around the Néel temperature, $T_N = 190$ K (details are shown in Section S3, Supporting Information). The observation that $T_B$ is lower than $T_N$ is consistent with previous studies.[13]

As shown in Figure 2b, these films are then patterned into Hall bar devices to detect the torque efficiency (details are in the Experimental Section). At 300 K, the spin torque efficiency ($\theta_{ST}$) of c-$Al_2O_3$/Hf (8 nm)/Py is estimated to be −0.22, aligning with the negative spin Hall angle of −0.2 observed for Hf layer thicker than 6 nm.[14] Figure 2c compares the absolute value of $\theta_{ST}$ for $c$-Al$_2$O$_3$/Hf (8 nm)/Py and $c$-Al$_2$O$_3$/Hf (8 nm)/Cr$_2$O$_3$ (20 nm)/Py. For Hf/Py, $|\theta_{ST}|$ remains stable as temperature decreases from 300 to 40 K, which is consistent with the behavior observed in other heavy metals.[15] However, the insertion of a 20 nm Cr$_2$O$_3$ layer, reduces $|\theta_{ST}|$ to 0.013 at 300 K, which is ascribed to the blocking of the electron-mediated spin currents by the paramagnetic Cr$_2$O$_3$ layer. As temperature decreases to around 170 K, $|\theta_{ST}|$ increases to 0.069 when the Cr$_2$O$_3$ layer transits from paramagnetic to AFM order. The amplitude of 0.069 is much smaller than that of 0.22 in Hf/Py, indicating that strong magnon torques cannot be induced when **n** is perpendicular to $\sigma_y$. These results are consistent with the magnon transmission in the YIG/Cr$_2$O$_3$ (0001)/Pt system.[10]

In contrast, the behavior of magnon torques on the $a$-Al$_2$O$_3$ substrates is quite different. The $|\theta_{ST}|$ for $a$-Al$_2$O$_3$/Hf/Py is similar to that of $c$-Al$_2$O$_3$/Hf/Py. However, for $a$-Al$_2$O$_3$/Hf (8 nm)/Cr$_2$O$_3$ (20 nm)/Py, $|\theta_{ST}|$ increases significantly from 0.016 at 300 K to 0.134 at 160 K. This gradual increase of $|\theta_{ST}|$ is consistent with the decrease in the SMR ratio of Hf/Cr$_2$O$_3$ (details are shown in Section S3, Supporting Information), indicating the progressive phase transition from the paramagnetic to the antiferromagnetic phase. We observe the $\theta_{ST}$ of −0.02 for the Cr$_2$O$_3$/Py interface on both $c$-Al$_2$O$_3$ and $a$-Al$_2$O$_3$ substrates, ruling out substantial contributions from the interface itself (details are shown in Section S4, Supporting Information). Therefore, this large $|\theta_{ST}|$ is attributed to magnon torques. The observation of magnon torques in the $a$-Al$_2$O$_3$/Hf/Cr$_2$O$_3$/Py system, but not in the $c$-Al$_2$O$_3$/Hf/Cr$_2$O$_3$/Py system, indicates that magnons are generated only when $\sigma_y$ in Hf aligns parallel to the **n** of Cr$_2$O$_3$.[16] This crystalline dependence of magnon torques is consistent with previous observations of magnon transmission in YIG/Cr$_2$O$_3$ (11$\bar{2}$0)/Pt system.[17] Moreover, the temperature ($T_p$) at which $|\theta_{ST}|$ shows a peak is close to $T_B$ (170 K). This is consistent with the observations in other AFM materials, where the temperature of maximum magnon transmission is comparable to the Néel temperature.[18]

In addition, we conduct in-plane second harmonic Hall measurements with varying the thickness ($t$) in Hf (8 nm)/Cr$_2$O$_3$ ($t$)/Py, summarized in Figure 3. At 300 K, on both $c$-Al$_2$O$_3$ and $a$-Al$_2$O$_3$ substrates, the data show a noticeable decline with the insertion of Cr$_2$O$_3$ layer and remain below 0.02 with an increase of $t$. This behavior is attributed to the paramagnetic Cr$_2$O$_3$ layers at 300 K, effectively blocking the electron-mediated spin currents. However, at 200 K, the behavior of $a$-Al$_2$O$_3$/Hf/Cr$_2$O$_3$/Py is different. The $|\theta_{ST}|$ gradually rises as $t$ increases beyond 5 nm, reaching a peak of 0.108 at $t$ = 20 nm. This peak is a signature of magnon torques, indicating the formation of AFM order within the Cr$_2$O$_3$ layer.[7, 8b, 9] To quantify the torque efficiency loss caused by the insertion of Cr$_2$O$_3$, we define the torque efficiency ratio as $r_\theta = \frac{|\theta_{ST}|_{peak}}{|\theta_{ST}|_{t=0}}$, where $|\theta_{ST}|_{t=0}$ represents the spin torque efficiency of Hf/Py, and $|\theta_{ST}|_{peak}$ represents the peak magnon torque efficiency of Hf/Cr$_2$O$_3$/Py. At 200 K, $r_\theta$ is determined to be 50.2 %, which matches values observed in Bi$_2$Se$_3$/NiO/Py,[7] Bi$_2$Te$_3$/NiO/Py,[8b] and WTe$_2$/NiO/Py[9] systems. At 100 K, the $|\theta_{ST}|$ peak shifts to $t = 15$ nm, with a maximum value of 0.095. This shift suggests that AFM order emerges in thinner Cr$_2$O$_3$ layers at lower temperatures.[10] Moreover, $r_\theta$ decreases to 43.4 %, indicating a suppression of magnon torques at lower temperatures. This behavior aligns with previous observations that magnons are quenched as temperature decreases.[8a] For c-Al$_2$O$_3$/Hf/Cr$_2$O$_3$/Py, weak peaks are observed at 200 K and 100 K as $t$ increases. The values of $r_\theta$ at 200 K and 100 K are 31.8 % and 22.8 %, respectively. These values are lower than those for a-Al$_2$O$_3$/Hf/Cr$_2$O$_3$/Py, indicating a larger torque efficiency loss in Cr$_2$O$_3$ with out-of-plane **n**. This difference confirms that magnon torques are suppressed in AFM Cr$_2$O$_3$ layers, where **n** is perpendicular to $\sigma_y$.

At 200 K, the $|\theta_{ST}|$ of a-Al$_2$O$_3$/Hf/Cr$_2$O$_3$/Py experiences an exponential decay for $t \geq 20$ nm, primarily due to dominant spin angular momentum loss of magnons via magnon-phonon and magnon-magnon interactions. We consider the magnon propagation in the Cr$_2$O$_3$ layer to be diffusive, so this decay can be fitted using an exponential decay function $\theta_{ST} = \theta_p \exp\left(\frac{-(t-t_p)}{l_m}\right)$,[8b, 19] where $t_p$ is the Cr$_2$O$_3$ thickness of Hf/Cr$_2$O$_3$/Py at which the magnon torque efficiency reaches its peak, $\theta_p$ is the corresponding peak magnon torque efficiency, and $l_m$ represents the magnon diffusion length. At 200 K, $l_m$ is determined to be 28.8 nm, which is comparable to values reported for NiO based magnon torque systems, such as Bi$_2$Se$_3$/NiO/Py and Bi$_2$Te$_3$/NiO/Py.[7, 8b] Furthermore, $l_m$ decreases to 22.9 nm at 100 K. It has been demonstrated that the magnon diffusion length for Y$_3$Fe$_5$O$_{12}$/Pt[20] and Tm$_3$Fe$_5$O$_{12}$/Pt[21] decreases with decreasing temperature, similar to what we observe in our Hf/Cr$_2$O$_3$ system. This decrease in $l_m$ at lower temperatures is attributed to the enhanced anisotropy of Cr$_2$O$_3$ as temperature decreases near $T_N$.[22] Therefore, both temperature and thickness dependence of $|\theta_{ST}|$ confirm the strong influence of the crystalline structure of Cr$_2$O$_3$ on magnon torques in the Hf/Cr$_2$O$_3$/FM system.

We subsequently utilize magnon torques to achieve magnetization switching of a FM layer with PMA. Ti (2 nm)/Co$_{20}$Fe$_{60}$B$_{20}$ (1 nm)/MgO (2 nm)/TaO$_x$ (1.5 nm) layers are deposited on top of a-Al$_2$O$_3$/Hf (8 nm), c-Al$_2$O$_3$/Hf (8 nm)/Cr$_2$O$_3$ (20 nm), and a-Al$_2$O$_3$/Hf (8 nm)/Cr$_2$O$_3$ (20 nm), using magnetron sputtering (details are shown in the Experimental Section). Ti is selected as the buffer layer due to its long spin diffusion length and its ability to enhance the perpendicular magnetization of CoFeB.[23] These films are patterned into Hall bar devices with a width of 5 μm and a length of 10 μm. Since the magnon torque efficiency of a-Al$_2$O$_3$/Hf (8 nm)/Cr$_2$O$_3$ (20 nm)/Py peaks at 160 K (Figure 2d), switching measurements are performed at 160 K.

The square-shaped anomalous Hall loops in Figure 4a confirm the PMA of CoFeB layer. Notably, the coercivity fields of these films are similar, with the difference of only 5%, allowing for the use of critical switching current density ($J_c$) to compare the switching efficiency. Moreover, no exchange bias is observed across the temperature range of 300 to 40 K (details are shown in Section S5, Supporting Information), which is attributed to the separation between the Cr$_2$O$_3$ and CoFeB layers by the Ti layer.[8b] Switching measurements are performed under different in-plane magnetic fields by injecting current pulses with a duration of 100 μs and varying amplitudes to the Hall bar devices. After each pulse, the Hall voltage is probed using a small direct current of 0.1 mA.

As shown in Figure 4b, a clear switching window is observed for a-Al$_2$O$_3$/Hf/CoFeB under an in-plane magnetic field ($\mu_0 H_x$) of ±10 mT, with a switching ratio of 100% (defined as the ratio of switching resistance to the anomalous Hall resistance). The switching is anticlockwise at +10 mT and clockwise at −10 mT, which is similar to typical PMA switching induced by a negative spin Hall material, such as Ta.[1a] The $J_c$ in Hf at 160 K is 2.45 × 10$^7$ A/cm$^2$, as calculated using the parallel resistance model (details are shown in Section S7, Supporting Information). This value is higher than the $J_c$ of 1.82 × 10$^7$ A/cm$^2$ measured at 300 K (details are shown in Section S9 and Section S10, Supporting Information). The increase in $J_c$ at lower temperatures is attributed to the enhanced anisotropy of the CoFeB layer.[15]

For c-Al$_2$O$_3$/Hf/Cr$_2$O$_3$/CoFeB, no switching behavior is observed, corresponding to the weak torque efficiency when **n** is perpendicular to $\sigma_y$. In contrast, for a-Al$_2$O$_3$/Hf/Cr$_2$O$_3$/CoFeB, a clear switching window is observed when the device channel is along the crystalline [1$\bar{1}$00] axis (**n** is parallel to $\sigma_y$), with $J_c$ of 4.09 × 10$^7$ A/cm$^2$. This demonstrates that the CoFeB layer can be switched by magnon torques. However, the switching ratio declines to 80%, indicating the torque efficiency loss due to the insertion of Cr$_2$O$_3$. Furthermore, the switching ratio drops significantly when the device deviates from the [1$\bar{1}$00] axis (details are shown in Section S11, Supporting Information), further confirming that magnon torques are highly dependent on the relative orientation between **n** and $\sigma_y$. This angular-dependent switching behavior is unique to the Hf/Cr$_2$O$_3$/CoFeB, which can be attributed to the fact that Cr$_2$O$_3$ has an easy-axis anisotropy.

The power consumption density is determined by $\frac{P}{S} = \frac{1}{LW}\left(I_{FM}^2 \rho_{FM} \frac{L}{W t_{FM}} + I_{Hf}^2 \rho_{Hf} \frac{L}{W t_{Hf}}\right)$, where $S$ is the device area, and $L$ and $W$ are the length and width of the device.[9] $\rho_{FM}$ and $\rho_{Hf}$ are the resistivities of the CoFeB and Hf layers, respectively. $t_{FM}$ and $t_{Hf}$ are the thicknesses of CoFeB and Hf layers, respectively. $I_{FM}$ and $I_{Hf}$ are the respective currents flowing through the CoFeB and Hf layers. The power consumption for $a$-$Al_2O_3$/Hf/$Cr_2O_3$/CoFeB is estimated to be 1.19 mW·μm$^{-2}$, which is lower than that of $Bi_2Te_3$/NiO/CoFeB (2.02 mW·μm$^{-2}$, with an in-plane magnetic field of 10 mT).[8b] When the temperature increases to 300 K, the switching behavior of $a$-$Al_2O_3$/Hf/$Cr_2O_3$/CoFeB disappears (details are shown in Section S10, Supporting Information), corresponding to the weak $|\theta_{ST}|$ value of 0.016 at 300 K (Figure 2d). The switching behavior is observed only below the Néel temperature of $Cr_2O_3$, confirming its magnon-driven origin.

Figure 4c shows the switching phase diagram (detailed switching behaviors are shown in Figure S14, Supporting Information). For Hf/CoFeB, the $J_c$ slightly decreases with increasing the in-plane magnetic field, which is a typical behavior for electron mediated switching.[1a] A similar trend is observed for the $J_c$ of Hf/$Cr_2O_3$/CoFeB, which is consistent with NiO based magnon mediated switching systems.[8b, 9] With the insertion of $Cr_2O_3$ between Hf and CoFeB layers, $J_c$ increases 1.6 times compared to that of Hf/CoFeB. In comparison, $J_c$ for $Bi_2Te_3$/NiO/CoFeB is more than twice that of $Bi_2Te_3$/CoFeB.[8b] This indicates that the insertion of crystalline $Cr_2O_3$ with in-plane **n** cause less spin angular momentum loss than the use of polycrystalline NiO.

## 3. Conclusion

We have observed magnon torques in all-sputtered Hf/$Cr_2O_3$/FM heterostructures, which are strongly dependent on the crystalline structure of antiferromagnetic $Cr_2O_3$. The magnon torque can be generated only when the Néel vector of $Cr_2O_3$ is parallel to the spin polarization in Hf, while it is suppressed when the Néel vector is perpendicular to the spin polarization. Especially, we realize magnon-torque-driven switching of FM layer with PMA. At 160 K, the critical switching current density is measured at $4.09 \times 10^7$ A/cm$^2$, with a corresponding magnon torque efficiency of −0.134. The power consumption density of our device is 1.19 mW·μm$^{-2}$, which is lower than that of $Bi_2Te_3$/NiO/FM system. In our experiments, magnon torques are excited by injecting charge currents into the Hf layer, leading to a reduced torque efficiency compared to its electron-mediated counterpart, Hf/FM. This reduction is less than the polycrystalline NiO due to the ordered structure of $Cr_2O_3$, indicating that the magnon torque efficiency can be enhanced through material optimization. Furthermore, the use of sputtered Hf, instead of topological materials, allows for integration into an industry-compatible sputtering process. Therefore, the Hf/Cr$_2$O$_3$ system is a more effective magnon torque platform compared to the topological materials/NiO system. In the future, we anticipate the generation of magnon torques based on electric-field-induced magnons in magnetoelectric multiferroic materials,[24] eliminating charge currents and significantly reducing the power consumption. We note that the $T_N$ of Cr$_2$O$_3$ is currently 190 K, which is insufficient for room-temperature applications. However, several strategies, such as epitaxial or chemical strain engineering, present promising pathways to increase $T_N$ to 400 K,[25] making Cr$_2$O$_3$ relevant for industrial applications. In addition, the Néel vector of boron (B) doped Cr$_2$O$_3$ can be manipulated using a gate voltage,[26] enabling the development of transistors for magnon torques. Our findings highlight the dependence of magnon torques on Néel vector orientation of antiferromagnetic insulators, broadening the potential applications of magnon torques.

## 4. Experimental Section/Methods

*Sample Preparation and Magnetic Properties Characterization*: Hf and Cr$_2$O$_3$ films were grown by magnetron sputtering onto Al$_2$O$_3$ (0001) and Al$_2$O$_3$ (11$\bar{2}$0) substrates in an ultra-high vacuum (UHV) chamber with the base pressure of $4 \times 10^{-9}$ Torr. The deposition process was conducted using a Cr$_2$O$_3$ target at 400 °C under an argon pressure of 2 mTorr, with the deposition rate of 0.42 nm·min$^{-1}$. Before deposition, the substrates were *in situ* annealed at 600 °C for 1 hour. For in-plane second harmonic Hall measurements, a 6 nm thick Ni$_{81}$Fe$_{19}$ (Py) layer with in-plane magnetic anisotropy and a 3nm thick SiO$_2$ capping layer were deposited by magnetron sputtering, after cooling the substrate to room temperature. For switching measurements, Ti, Co$_{20}$Fe$_{60}$B$_{20}$, MgO, and TaO$_x$ capping layers were sequentially deposited at room temperature. After annealing at 200 °C for 30 min with an external out-of-plane magnetic field of 0.5 T, the film exhibited good PMA properties. The *M-H* curves were measured by a vibrating sample magnetometer (VSM).

*In-plane Second Harmonic Hall Measurements*: The films were patterned into Hall bar devices with a width of 5 μm and length of 50 μm using photolithography and ion milling techniques. Subsequently, electrodes were fabricated to interface with the devices, by photolithography, magnetron sputter, and lift-off techniques. The alternating current at a frequency of 13.7 Hz was applied using a Keithley 6221 source meter. Simultaneously, the first and second harmonic Hall signals were detected using a lock-in amplifier (Stanford SR830). An external magnetic field ($\mu_0 H_{ext}$) ranging from 0.2 to 1.25 T was applied in the film plane.

The samples were measured in a cryogenic chamber, allowing the temperature to be cooled from 300 to 40 K.

*Switching Measurements*: The films were patterned into Hall bar devices with a width of 5 μm and a length of 10 μm, and the electrical contact pad patterns were fabricated. Electrical current pulses with 100 μs duration were applied by a Keithley 6221 source meter. After each pulse, a small direct current of 0.1 mA was applied by the Keithley 6221 source meter, and the transverse voltage was measured by a Keithley 2182A nanovoltmeter.

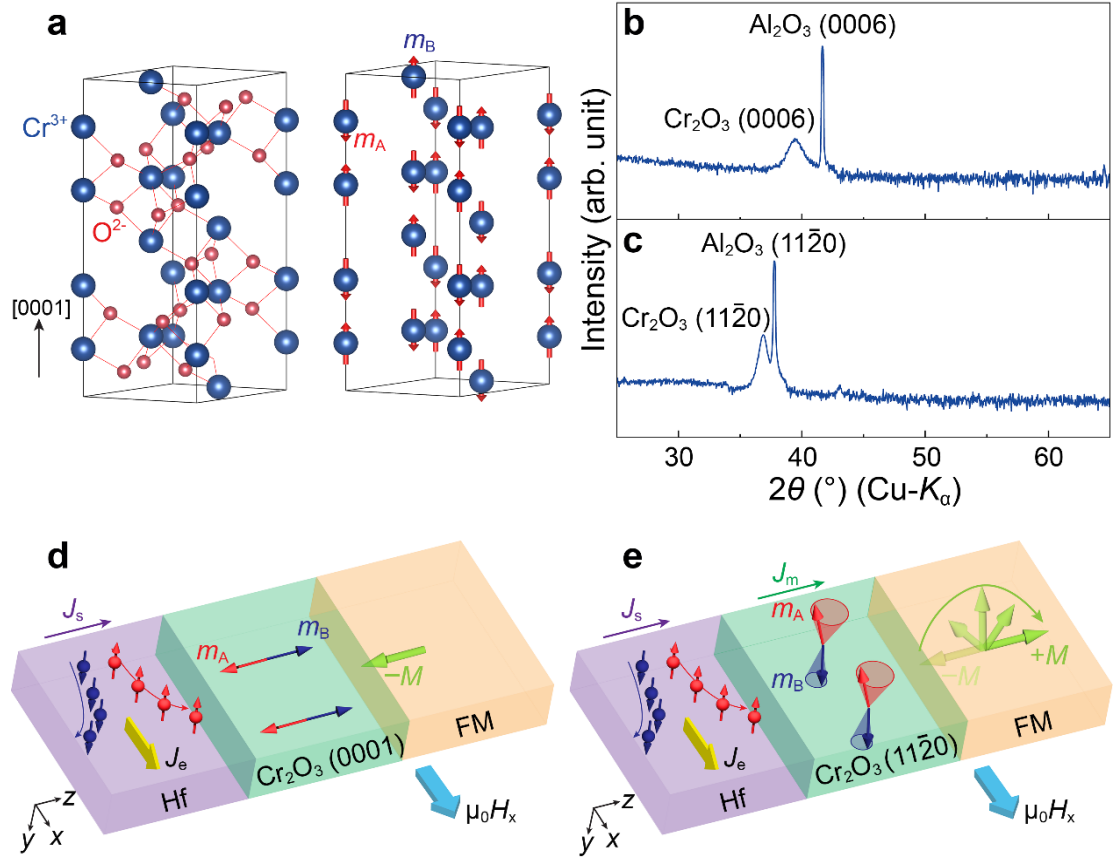

**Figure 1. Characterization of Hf/Cr$_2$O$_3$/FM heterostructures.** a) Atomic and magnetic structures of Cr$_2$O$_3$. The red arrows represent the magnetic moments of Cr$^{3+}$. The X-ray diffraction data for b) *c*-Al$_2$O$_3$/Hf (8 nm)/Cr$_2$O$_3$ (20 nm)/Py (6 nm) and c) *a*-Al$_2$O$_3$/Hf (8 nm)/Cr$_2$O$_3$ (20 nm)/Py (6 nm). The Hf peak is not observed due to the thin Hf layer. 20 nm thick Hf layers on *c*-Al$_2$O$_3$ and *a*-Al$_2$O$_3$ exhibit clear peaks, as detailed in Section S1, Supporting Information. d) Magnon torques are absent in the *c*-Al$_2$O$_3$/Hf/Cr$_2$O$_3$/FM heterostructures, where Néel vector of Cr$_2$O$_3$ is out-of-plane. $J_s$ represents the spin current. $J_e$ represents the electron current. e) Magnetization switching by magnon torques in the *a*-Al$_2$O$_3$/Hf/Cr$_2$O$_3$/FM heterostructures, where the Néel vector of Cr$_2$O$_3$ is in-plane. $J_m$ represents the magnon current. An in-plane magnetic field ($\mu_0 H_x$) is applied along the current direction.

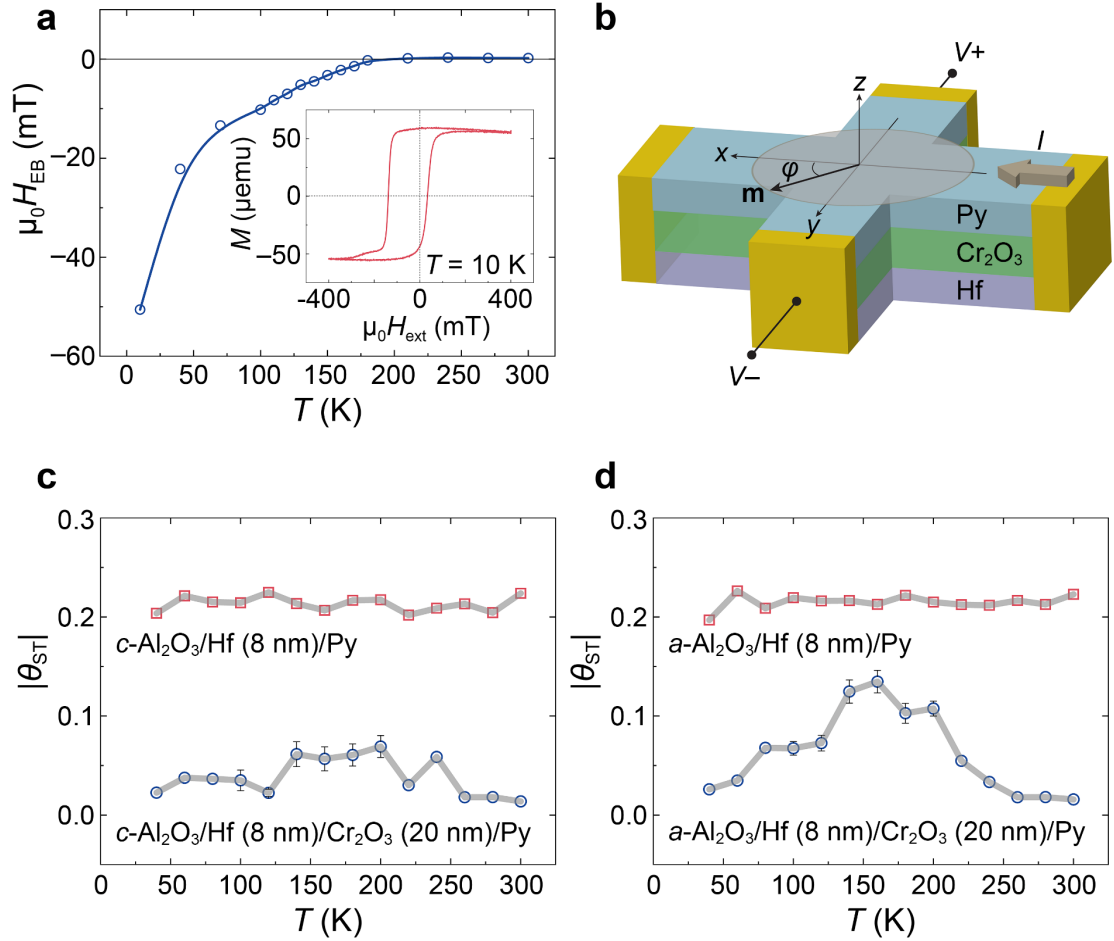

**Figure 2. Temperature dependence of torque efficiency ($\theta_{ST}$) for Hf/Cr$_2$O$_3$/Py heterostructures.** a) Dependence on temperature ($T$) of the exchange bias field ($\mu_0 H_{EB}$) of $a$-Al$_2$O$_3$/Hf (8 nm)/Cr$_2$O$_3$ (20 nm)/Py. The inset figure is typical exchange-biased $M$-$\mu_0 H_{ext}$ hysteresis loop at 10 K. b) The schematic of in-plane second harmonic Hall measurements. The device channels are oriented along the [11$\bar{2}$0] axis on $c$-Al$_2$O$_3$ substrates and along the [1$\bar{1}$00] axis on $a$-Al$_2$O$_3$ substrates. The absolute values of torque efficiency ($|\theta_{ST}|$) for Hf (8 nm)/Py and Hf (8 nm)/Cr$_2$O$_3$ (20 nm)/Py on c) $c$-Al$_2$O$_3$ and d) $a$-Al$_2$O$_3$ substrates.

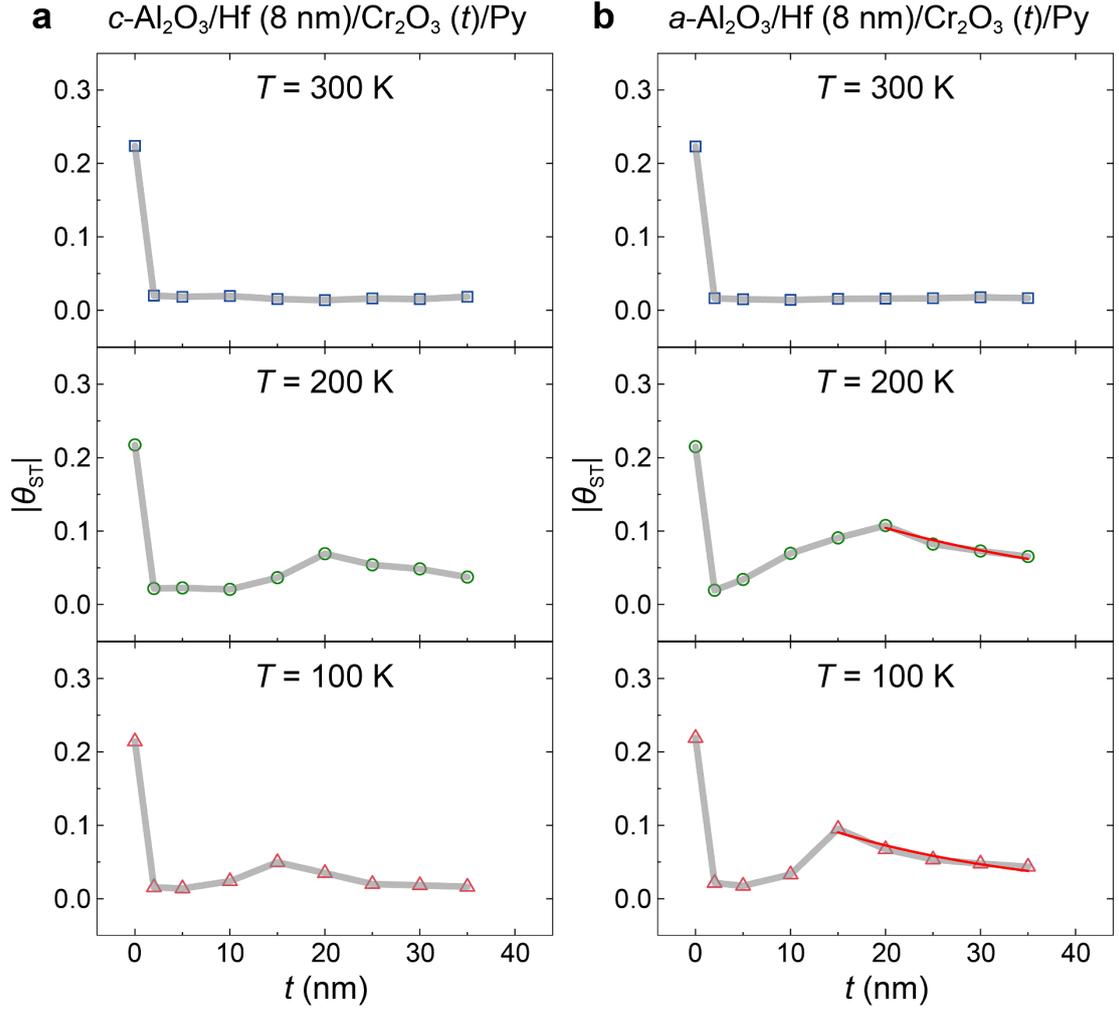

**Figure 3. Thickness dependence of torque efficiency ($\theta_{ST}$) for Hf/Cr$_2$O$_3$/Py heterostructures.** a) The absolute values of torque efficiency ($|\theta_{ST}|$) of c-Al$_2$O$_3$/Hf/Cr$_2$O$_3$/Py versus Cr$_2$O$_3$ thickness ($t$) at 300, 200, and 100 K. b) The absolute values of torque efficiency ($|\theta_{ST}|$) of a-Al$_2$O$_3$/Hf/Cr$_2$O$_3$/Py versus Cr$_2$O$_3$ thickness ($t$) at 300, 200, and 100 K. The red lines are exponential decay fittings of $|\theta_{ST}|$.

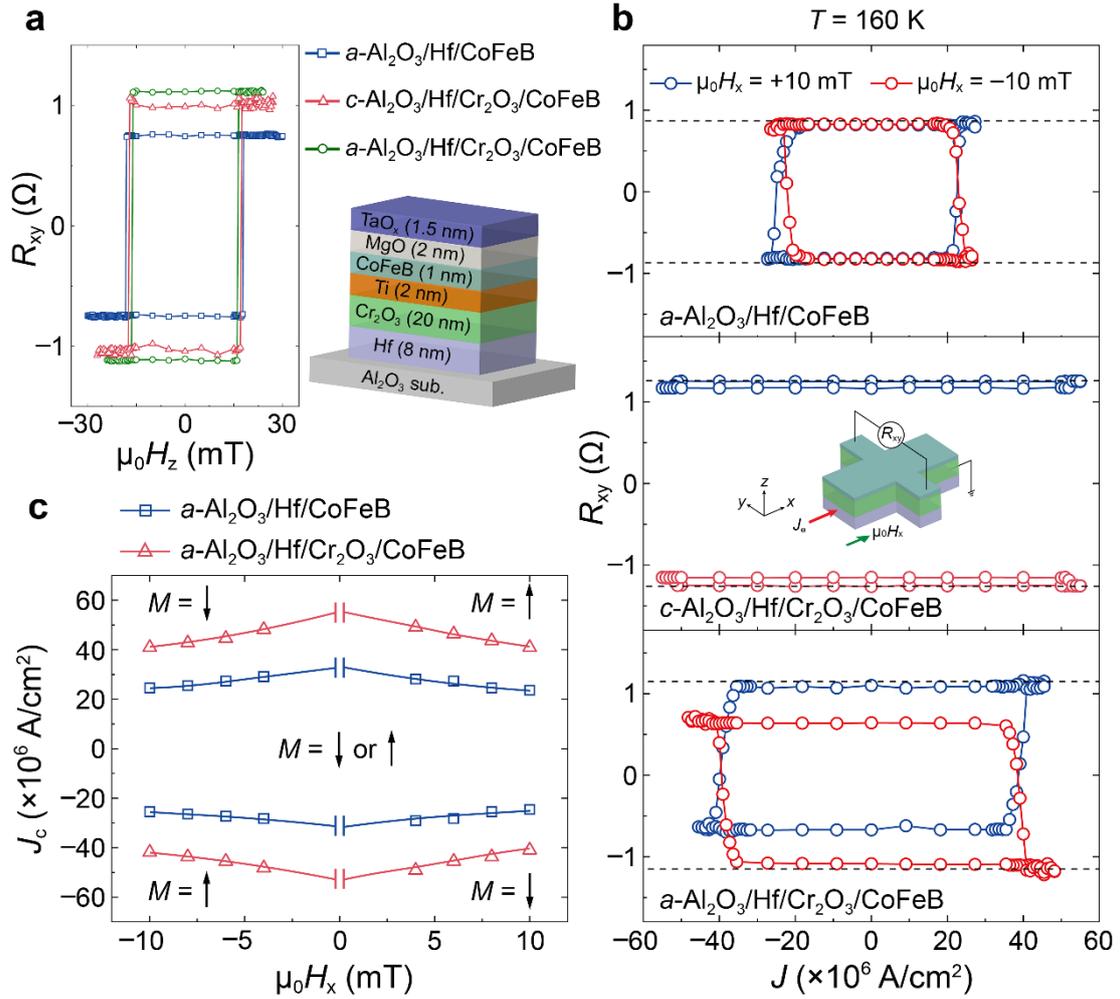

**Figure 4. Magnon-torque-driven magnetization switching in the Hf (8 nm)/Cr$_2$O$_3$ (20 nm)/CoFeB sample.** a) Schematic layout for Hf/Cr$_2$O$_3$/CoFeB stack and anomalous Hall curve measured by sweeping the magnetic field ($\mu_0H_z$) along the $z$ direction at 300 K. b) The magnetization switching behaviors for $a$-Al$_2$O$_3$/Hf/CoFeB, $c$-Al$_2$O$_3$/Hf/Cr$_2$O$_3$/CoFeB, and $a$-Al$_2$O$_3$/Hf/Cr$_2$O$_3$/CoFeB under in-plane magnetic field ($\mu_0H_x$) of −10 and +10 mT. The measurements are conducted at 160 K. The black dashed lines represent the anomalous Hall resistance. The inset is the schematic illustration of the measurement setup. The device channels are oriented along the [11$\bar{2}$0] axis on $c$-Al$_2$O$_3$ substrates and along the [1$\bar{1}$00] axis on $a$-Al$_2$O$_3$ substrates. c) $\mu_0H_x$ dependence of the critical switching current density ($J_c$).